\documentclass[twocolumn,apl,superscriptaddress]{revtex4-1}

\usepackage{graphicx}
\usepackage{dcolumn}
\usepackage{bm}
\usepackage{hyperref}
\usepackage{amsmath}
\usepackage{amssymb}
\usepackage[normalem]{ulem}
\usepackage[usenames,dvipsnames,svgnames,table]{xcolor}

\begin{document}

\preprint{APS/123-QED}

\title{Interface-induced heavy-hole/light-hole splitting of acceptors in silicon}

\author{J. A. Mol}
\affiliation{Centre for Quantum Computation and Communication Technology, School of Physics, University of New South Wales, Sydney, NSW 2052, Australia}
\affiliation{Department of Materials, University of Oxford, 16 Parks Road, Oxford OX1 3PH, UK}
\author{J. Salfi}
\affiliation{Centre for Quantum Computation and Communication Technology, School of Physics, University of New South Wales, Sydney, NSW 2052, Australia}
\author{R. Rahman}
\affiliation{Purdue University, West Lafayette, IN 47906, USA}
\author{Y. Hsueh}
\affiliation{Purdue University, West Lafayette, IN 47906, USA}
\author{J. A. Miwa}
\affiliation{Centre for Quantum Computation and Communication Technology, School of Physics, University of New South Wales, Sydney, NSW 2052, Australia}
\affiliation{Department of Physics and Astronomy, Interdisciplinary Nanoscience Center (iNANO), Aarhus University, 8000 Aarhus C, Denmark}
\author{G. Klimeck}
\affiliation{Purdue University, West Lafayette, IN 47906, USA}
\author{M. Y. Simmons}
\affiliation{Centre for Quantum Computation and Communication Technology, School of Physics, University of New South Wales, Sydney, NSW 2052, Australia}
\author{S. Rogge$^\ast$}
\affiliation{Centre for Quantum Computation and Communication Technology, School of Physics, University of New South Wales, Sydney, NSW 2052, Australia}

\collaboration{$^\ast$To whom correspondence should be addressed; E-mail:  s.rogge@unsw.edu.au.}

\date{\today}
             
\begin{abstract}
The energy spectrum of spin-orbit coupled states of individual sub-surface boron acceptor dopants in silicon have been investigated using scanning tunneling spectroscopy (STS) at cryogenic temperatures. The spatially resolved tunnel spectra show two resonances which we ascribe to the heavy- and light-hole Kramers doublets. This type of broken degeneracy has recently been argued to be advantageous for the lifetime of acceptor-based qubits [\textit{Phys. Rev. B} \textbf{88} 064308 (2013)]. The depth dependent energy splitting between the heavy- and light-hole Kramers doublets is consistent with tight binding calculations, and is in excess of 1 meV for all acceptors within the experimentally accessible depth range ($< 2$ nm from the surface). These results will aid the development of tunable acceptor-based qubits in silicon with long coherence times and the possibility for electrical manipulation.
\end{abstract}

\pacs{Valid PACS appear here}

\maketitle

Dopant atoms in silicon are attractive candidates for spin-based quantum computation. Recent studies have demonstrated long coherence-times for both ensembles of bulk donors \cite{Saeedi:2013dy} and individual  donors \cite{Muhonen:tb} in silicon. Meanwhile, rapid progress in scanning tunnelling microscopy (STM) based lithography has paved the way towards atomically precise placement of dopant atoms \cite{Fuechsle:2012bl}. While phosphorous donors in silicon remains among the most compelling candidates for dopant-based quantum computing to date, other impurity systems have recently drawn considerable attention. In particular, boron acceptors could provide a pathway towards electrically addressable spin-qubits via spin-orbit coupling \cite{Ruskov:2013kq} analogues to electrically driven spin manipulation in gate defined electron \cite{Nowack:2007du,NadjPerge:2010kw} and hole \cite{Pribiag:2013if} quantum dots in III-V materials. Compared with other spin-orbit qubits, acceptors in silicon have several advantages: gates are not required for hole confinement and each qubit experiences the same confinement potential; furthermore the hole-spin decoherence, due to the nuclear spin bath, can be effectively eliminated by isotope purification of the silicon host.

Unlike the ground state of donor-bound electrons in silicon, the acceptor-bound hole ground state is four-fold degenerate, reflecting the heavy-hole/light-hole degeneracy of the silicon valence band. Recent theoretical work has suggested the regime of long lifetimes for acceptors with a four-fold degenerate ground state is only accessible for small magnetic fields \cite{Ruskov:2013kq}. Interestingly, this work also suggests that symmetry breaking due to strain or electric fields could yield longer-lived qubits based on acceptor-bound holes at higher magnetic fields. The symmetry breaking perturbation of biaxial strain \cite{Ruskov:2013kq} renders the lowest two (qubit) levels within a Kramers degenerate pair, such that they do not directly couple to electric fields. Quantum confinement could provide a similar form of protection. In this Letter we demonstrate that the symmetry-reduction of a potential boundary renders the lowest two levels Kramers degenerate and heavy-hole like. Recent transport spectroscopy studies of an individual acceptor embedded in nano-scale transistors have shown that for acceptors $\sim$10 ~nm away from an interface the bulk-like four-fold degeneracy is maintained \cite{vanderHeijden:2014fp}. Here we demonstrate that the presence of a nearby interface ($<$2~nm) lifts the four-fold degeneracy of the acceptor-bound hole ground state by investigating the energy spectrum and wavefunctions of individual sub-surface boron acceptors using scanning tunnelling spectroscopy (STS). 

Recent low-temperature STM/STS studies have successfully probed the energy spectrum and wavefunctions of individual impurity atoms in Si \cite{Mol:2013dj,Sinthiptharakoon:2013il,Salfi:2014ka} and GaAs \cite{Yakunin:2007dv,Wijnheijmer:2009il}. These studies have revealed the influence of the semiconductor/vacuum interface on the ionization energy of sub-surface dopants \cite{Wijnheijmer:2009il,Mol:2013dj}, the spatially resolved structure of the dopant wavefunction \cite{Yakunin:2007dv,Sinthiptharakoon:2013il,Salfi:2014ka} and the mechanisms for charge-transport through these dopants \cite{Mol:2013dj,Miwa:2013bf,Salfi:2014ka}. Here, we use STS to measure the energy difference between the heavy-hole and light-hole states of individual B acceptors less than 2~nm away from the surface. The sample is prepared by repeated flash annealing a $8\times10^{18}$~cm$^{-3}$ boron doped Si(100) substrate in ultra-high vacuum. The flash annealing not only provides an atomically flat surface but also yields a region of low doping at the interface due to out-diffusion of the B impurities. As a result, the near-interface acceptors are weakly coupled to the bulk acceptors, which form a valence impurity band (VIB). Together with the STM tip, the sub-surface acceptors can be considered as a double-barrier system for single-hole tunneling \cite{Mol:2013dj} (see Fig.1(a)). As illustrated in the inset of Figure 1(b), individual acceptors are identified as protrusions at $V_b=-1.5$~V due to enhancement of the valence band density of states \cite{Mol:2013dj}.

\begin{figure}
 \includegraphics{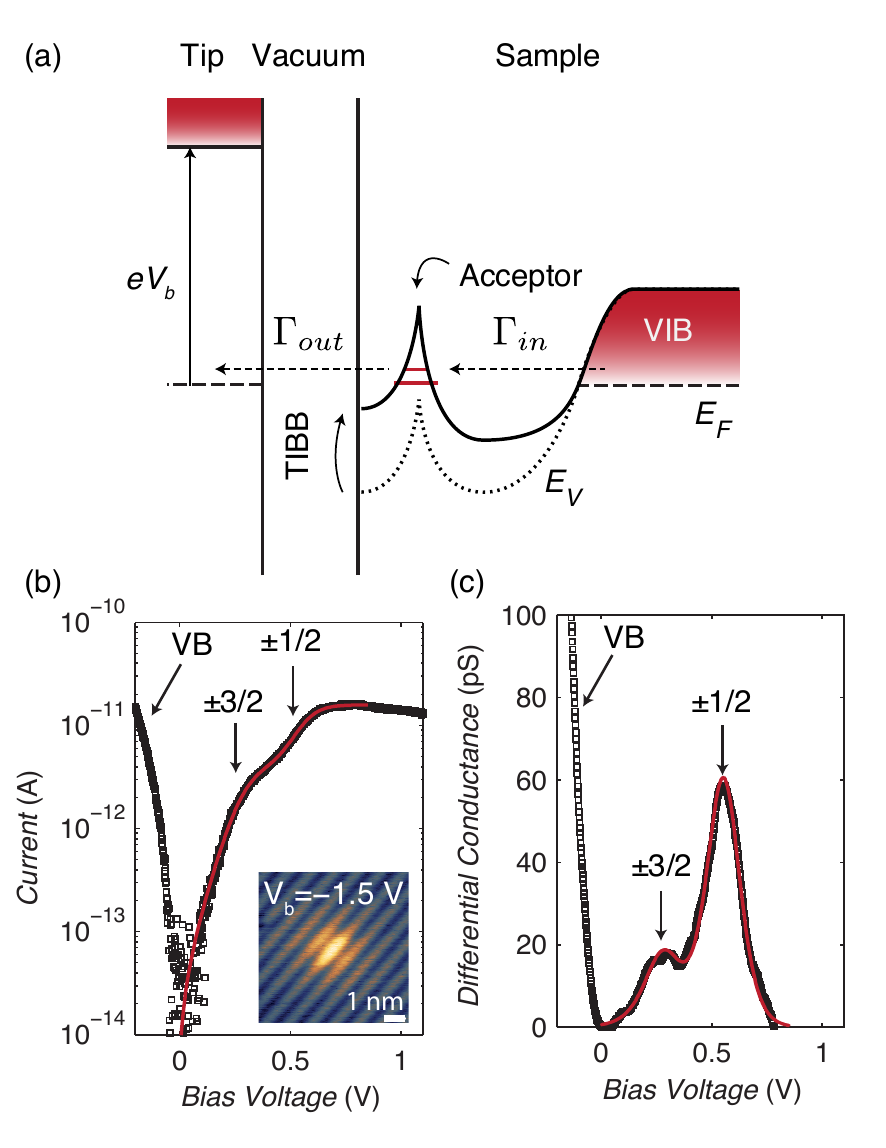}
  \caption{(a) Schematic energy diagram of the tunneling process. Applying a positive sample bias ($eV_b$) results in an upward tip-induced band bending (TIBB). When the energy of a state of an isolated acceptor near the interface is pulled above the Fermi level ($E_F$) there is a peak in the conductance due to resonant tunneling from the valence impurity band (VIB) through the acceptor state into the tip. In this schematic description, the top layer of silicon is depleted of dopants. The measured acceptors are well-isolated from each other, as well as the underlying nominally $8\times10^{18}$ cm$^{-3}$ doped region, denoted by VIB. (b) Current and (c) differential conductance measured directly above the isolated acceptor (open circles) as a function of bias voltage. The differential conductance within the band gap is fitted to the sum (solid line) of two thermally broadened Lorentzian line-shapes corresponding to resonant tunnelling through the lowest two acceptor states.}
  \label{fig1}
\end{figure}

Figure 1(b,c) shows the current and differential conductance measured over a single sub-surface acceptor. For the bias voltage $V_b<0$~V the charge transport is dominated by holes tunnelling directly from the Si valence band to the tip. Similarly, for $V_b>1.1$~V transport is dominated by holes tunnelling from the tip to the Si conduction band. The observed features in the band-gap, i.e. for $0$~V$<V_b<1.1$~V, can be attributed to holes tunnelling from the VIB through the localized acceptor states to the tip. Upon increasing the bias voltage from 0~V to 1.1~V the energy levels of the subsurface acceptor states shift due to tip induced band bending (TIBB). Each time an acceptor level enters the bias window, which is defined by the Fermi level of the tip and the Si substrate, this opens an additional channel for transport resulting in a stepwise increase in the current (Fig. 1(b)) and a corresponding peak in the differential conductance as shown in Fig. 1(c). The conductance peaks are only observed in the presence of an acceptor, and for each measured acceptor we observe two conductance peaks in the band-gap which we attribute to the lowest two acceptor states entering the bias window.

\begin{figure}
 \includegraphics{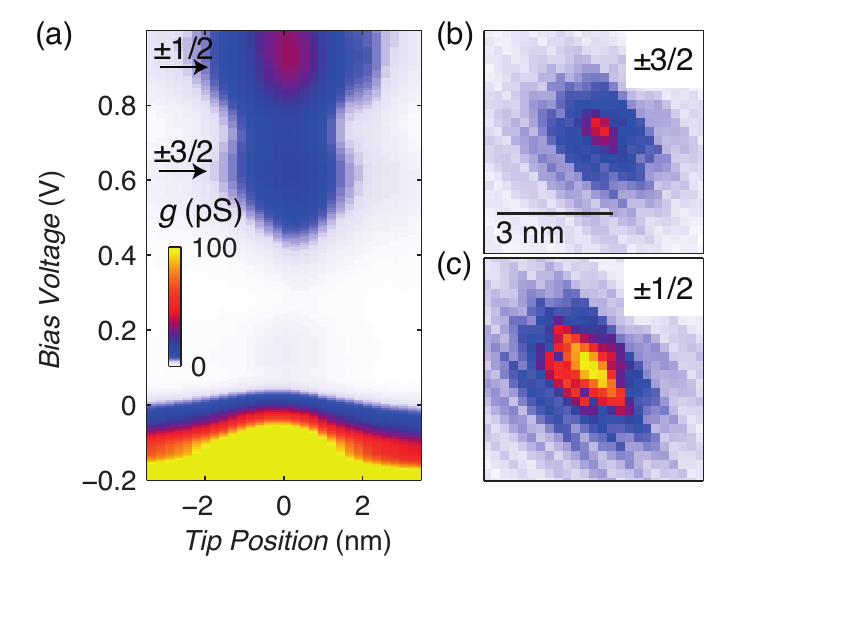}
  \caption{(a) Spatially resolved differential conductance map as a function of tip position and bias voltage measured over a sub-surface acceptor (different from Fig. 1). (b,c) Probability density map of the lowest two acceptor states.}
  \label{fig2}
\end{figure}

A spatially resolved differential conductance map is shown in Fig. 2(a). The amplitude of the differential conductance measured over the acceptor is a measure of the probability density of the impurity wavefunction (Fig. 2(b,c)). The probability density maps of the lowest two acceptor states have the same spatial extent and symmetry, i.e. neither state shows anti-nodes in the surface plane. The similarity between the probability density map of the two conductance peaks is consistent with what is expected for $\pm 3/2$ and $\pm 1/2$ states.  Both are expected to have the observed s-like envelopes, with similar spatial extents \cite{Schechter:1962kw}. In contrast, a two-hole state would have a larger spatial extent. When two states have the same envelope wavefunctions the energy difference between these two states can only arise from the difference in pseudo-spin, i.e. from non-degenerate heavy-hole and light-hole states. In the remainder of this Letter we will discuss how the perturbation of the subsurface acceptor wavefunctions by an interface lifts the degeneracy between the heavy- and light-hole Kramers doublets that make up the four-fold degenerate ground state of an unperturbed (bulk) acceptor.

Based on the tunneling spectra of the single acceptor we can now extract the energy splittings of the states, and show that they are in reasonable agreement with tight binding predictions for the s-like 3/2 and 1/2 states, and too small to be associated with charging processes where a second hole is bound to the acceptor. If the lever arm $\alpha=dE_i/edV_i$ that couples the chemical potential to the bias voltage is known, the voltage $V_i$ for which a localized state $i$ is brought into resonance is a direct measure for the eigenenergy $E_i$ of this state. We determine $\alpha$ by fitting the conductance to a thermally broadened Lorentzian \cite{Foxman:1993jq} (see Fig. 1(b)):
\begin{equation}
\label{eq:conv}
\begin{split}
g(V) \propto \int^{+\infty}_{-\infty}&\cosh^{-2}(E/2k_BT) \\
&\times \frac{\frac{1}{2}\hbar\Gamma}{(\frac{1}{2}\hbar\Gamma)^2+(\alpha e[V-V_i] - E)^2}dE,
\end{split}
\end{equation}
where $k_B$ is the Boltzmann factor, $T$ the temperature (we assume $T=4.2$~K for the fit), $V_i$ the voltage corresponding to the center of the $i^{th}$ conductance peak, $\hbar$ Planck's constant and $\Gamma$ the sum of the tunnel-in and tunnel-out rates. The interface induced splitting is given by $\delta E=\alpha(V_{\pm1/2}-V_{\pm3/2})$. 

The distance of the subsurface acceptors to the interface is measured from the spectral shift of the valence band edge \cite{Mol:2013dj}:
 \begin{equation}\label{prx_eq:coulomb}
\Delta E_V \approx e\frac{e-Q}{4\pi\epsilon_{0}\epsilon_{Si}}\frac{1}{\sqrt{s^2+d^2}},
\end{equation}
where $d$ is the acceptor depth and $Q=e(\epsilon_{v}-\epsilon_{Si})/(\epsilon_v+\epsilon_{Si})$ is the image charge due to the mismatch between the dielectric constants $\epsilon_{v}$ and $\epsilon_{Si}$ of the vacuum and silicon, respectively. The onset voltage for tunneling from the valence band $V_V$ is determined by finding the voltage axis intercept of the linear extrapolation of the normalised conductance curve at its maximum slope point \cite{Feenstra:1994ur}. The spectral shift of the valence-band edge is fitted to equation~\ref{prx_eq:coulomb} using the depth $d$ of individual acceptors and the modified dielectric constant $Q$ as two, independent, fitting parameters. The obtained values for $Q$ for all measured acceptors agree within experimental error with the expected value $Q=e(\epsilon_{v}-\epsilon_{Si})/(\epsilon_v+\epsilon_{Si})$ following the classical half-space approach and experimental values that have previously been reported from STM experiments \cite{Teichmann:2008bh,Lee:2010ko}.

\begin{figure}
 \includegraphics{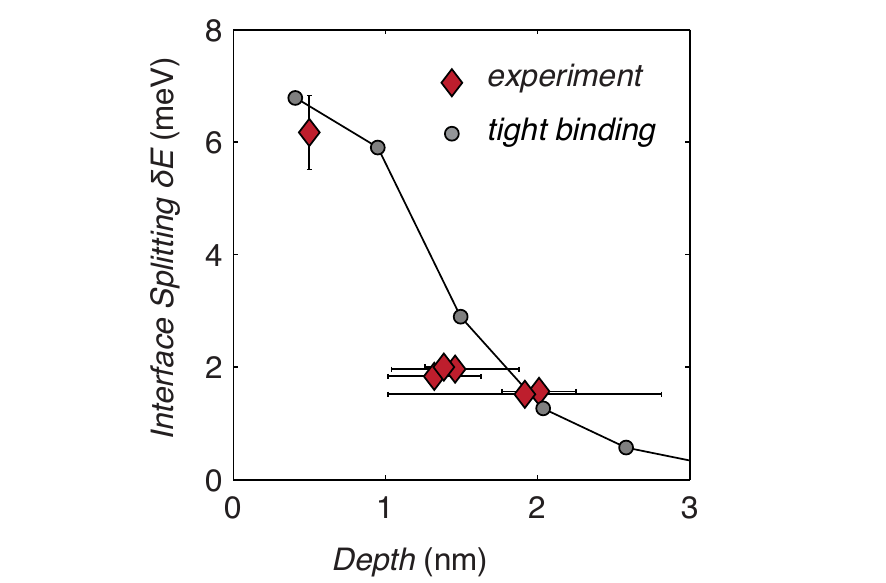}
  \caption{Experimental and theoretical depth dependence of the interface induced energy splitting. The vertical error bars on the energy-level splitting are proportional to the confidence of the fit of $\alpha$.  Errors in $V_i$ are negligible. The horizontal error bars are proportional to the confidence of the fit of $d$ which decreases for acceptors further from the interface as the spectral shift of the valence-band maximum becomes smaller.
}
  \label{fig3}
\end{figure}

Figure 3 shows the measured energy splitting $\delta E$ between the lowest two acceptor states as a function of acceptor depth for six different acceptors demarcated by the solid diamonds. These results are compared with a fully atomistic tight binding (TB) simulation marked by the grey circles. The tight-binding Hamiltonian of 1.4 million silicon atoms with a boron acceptor was represented with a 20-orbital sp3d5s* basis per atom including nearest-neighbor and spin-orbit interactions. An acceptor was represented by a Coulomb potential of a negative charge screened by the dielectric constant of Si and subjected to an onsite cutoff potential $U_0$ \cite{Rahman:2007ev}. The model provides an accurate solution for the single-hole eigenstates of a bulk acceptor, and an acceptor near an interface. The magnetic field is represented by a vector potential in a symmetric gauge and entered through a Peierls substitution. The full TB Hamiltonian is solved in NEMO3D \cite{Klimeck:2007gs} by a parallel Block Lanczos algorithm, and the relevant low energy acceptor wavefunctions are obtained. From the calculated magnetic field dependence of the acceptor states (not shown) we infer that the lowest Kramers doublet has a predominant heavy-hole ($\pm3/2$) character  whereas the second Kramers doublet has a predominant light-hole ($\pm1/2$) character. 

From the combination of the spatial and spectral measurements, we can conclude that the observed states belong to the s-like manifold which contains a $\pm 3/2$ Kramers doublet and a $\pm 1/2$ Kramers doublet.  Not only are both of the envelopes s-like, but the few meV energy scale of the splitting is too small to be associated with higher excited states (at 25 meV) \cite{Wright:1967ee} or charging transitions (47 meV) \cite{Burger:1984jn} of an acceptors.  We note that the charging energy near the surface could fall as low as 20 meV \cite{Salfi:2014ka}, but this is still too large for our results.

We observe an increase of $\delta E$ for acceptors closer to the interface, as predicted by the TB calculations. Importantly, all acceptors studied showed an energy splitting in excess of 1 meV. For comparison, the electric field required to obtain a splitting of 0.5 meV would be 40 MV/m \cite{Kopf:1992gj}, much greater than the electric field required for field ionization 5 MV/m \cite{Smit:2004fx}. Our results therefore demonstrate that the presence of an interface provides an effective way to energetically isolate a single Kramers doublet that could serve as the working levels of a spin-qubit. Such a qubit could benefit from elimination of the nuclear spin bath by isotope purification, and could provide a new route towards an electrically controllable spin qubit.   

\begin{acknowledgements}
This research was conducted by the Australian Research Council Centre of Excellence for Quantum Computation and Communication Technology (project number CE110001027) and the US National Security Agency and the US Army Research Office under contract number W911NF-08-1-0527. J.A.M. received funding from the Royal Society Newton International Fellowship scheme. M.Y.S. acknowledges an ARC Federation Fellowship. S.R acknowledges an ARC Future Fellowship and FP7 MULTI. The authors are grateful to D. Culcer for helpful discussions.
\end{acknowledgements}
%

\end{document}